\documentclass[superscriptaddress,amsmath,amssymb,floatfix,aps,prd,nofootinbib,reprint]{revtex4-1}
\usepackage{natbib}
\usepackage{graphics}
\usepackage{multirow}
\usepackage{amsbsy}
\usepackage{footmisc}
\usepackage{hyperref}
\usepackage{amsmath}
\usepackage{aas_macros}

\usepackage{array}
\usepackage[mathscr]{eucal}
\usepackage{morefloats}
\usepackage{amssymb}
\usepackage{txfonts}
\usepackage{placeins}
\usepackage{natbib}

\DeclareMathAlphabet{\mathbfsf}{\encodingdefault}{\sfdefault}{bx}{n}

\usepackage{verbatim}
\usepackage{graphicx}
\usepackage{epsf}
\usepackage{subfigure}
\usepackage{color}
\usepackage{threeparttable}
\usepackage{comment}
\usepackage{epsfig}
\usepackage{xspace}
\usepackage{enumitem}
\usepackage{hyperref}
\usepackage{ulem}
\usepackage{courier}
\usepackage{appendix}
\usepackage{comment}
\usepackage[usenames,dvipsnames,svgnames]{xcolor}
\usepackage{latexsym}
\usepackage{url}

\newcommand{\refsec}[1]{Sec.~\ref{sec:#1}}
\newcommand{\reffig}[1]{Fig.~\ref{fig:#1}}
\newcommand{\refeqn}[1]{Eq.~\ref{eqn:#1}}
\newcommand{\calibM}{\ensuremath{y}}
\newcommand{\calibC}{\ensuremath{c}}
\newcommand{\reftab}[1]{Tab.~\ref{tab:#1}}

\hypersetup{
  colorlinks = true,
  urlcolor = blue,
  linkcolor=blue,
  citecolor=blue
}

\usepackage[applemac]{inputenc}

\def\planck{\textit{Planck}}
\def\sptpol{{\sc SPTpol}}
\def\spt{{\sc SPT}}
\def\act{{\sc ACT}}
\def\wmap{{\sc WMAP}}
\def\pcal{$P_{\rm cal}$}
\def\tcal{$T_{\rm cal}$}
\newcommand{\mksym}[1]{\ifmmode {\rm #1}\else #1\fi}
\newcommand{\dataplus}{\allowbreak+}
\newcommand{\lowE}{\mksym{lowE}}

\newcommand{\TT}{\mksym{TT}}
\newcommand{\TE}{\mksym{TE}}
\newcommand{\EE}{\mksym{EE}}

\newcommand{\planckTEEE}{\planck \TE, \EE \dataplus\lowE}
\newcommand{\planckTTTEEE}{\planck \TT,\TE, \EE \dataplus\lowE}

\newcommand{\plik}{{\tt Plik}}
\newcommand{\commander}{{\tt Commander}}
\newcommand{\As}{A_{\rm s}}
\newcommand{\simall}{{\tt SimAll}}

\newcommand{\sroll}{\texttt{SRoll}}
\newcommand{\lcdm}{\texorpdfstring{{$\rm{\Lambda CDM}$}}{ΛCDM}}
\newcommand{\LCDM}{\lcdm}
\newcommand{\lnAs}{\ln(10^{10} A_{\rm s})}
\newcommand{\Alens}{A_{\rm L}}
\newcommand{\nnu}{N_{\rm eff}}
\newcommand{\mnu}{\sum m_{\nu}}
\begin{document}
\title{Breaking the degeneracy between polarization efficiency and cosmological parameters in CMB experiments}
\author{Silvia Galli}
\email{galli@iap.fr}
\affiliation{Sorbonne Universit\'{e}, CNRS, Institut d'Astrophysique de Paris, 98 bis Boulevard Arago, F-75014 Paris, France}

\author{W.~L.~Kimmy Wu}
\email{wlwu@slac.stanford.edu}
\affiliation{SLAC National Accelerator Laboratory \& KIPAC, 2575 Sand Hill Road, Menlo Park, CA 94025}
\affiliation{Kavli Institute for Cosmological Physics, University of Chicago, Chicago, Illinois 60637, U.S.A }

\author{Karim~Benabed}
\affiliation{Sorbonne Universit\'{e}, CNRS, Institut d'Astrophysique de Paris, 98 bis Boulevard Arago, F-75014 Paris, France}

\author{Fran\c{c}ois~Bouchet} 
\affiliation{Sorbonne Universit\'{e}, CNRS, Institut d'Astrophysique de Paris, 98 bis Boulevard Arago, F-75014 Paris, France}

\author{Thomas~M.~Crawford}
\affiliation{Kavli Institute for Cosmological Physics, University of Chicago, Chicago, Illinois 60637, U.S.A }
\affiliation{Department of Astronomy and Astrophysics, University of Chicago, 5640 South Ellis Avenue, Chicago, IL, 60637, USA}

\author{Eric~Hivon}
\affiliation{Sorbonne Universit\'{e}, CNRS, Institut d'Astrophysique de Paris, 98 bis Boulevard Arago, F-75014 Paris, France}

\date{\today}
\begin{abstract}
Accurate cosmological parameter estimates using polarization data of the cosmic microwave background (CMB) put stringent requirements on map calibration, as highlighted in the recent results from the \planck\ satellite. 
In this paper, we point out that a model-dependent determination of polarization calibration can be achieved by the joint fit of the \TE\ and \EE\ CMB power spectra. 
This provides a valuable cross-check to band-averaged polarization efficiency measurements determined using other approaches. 
We demonstrate that, in $\Lambda$CDM, the combination of the \TE\ and \EE\ constrain polarization calibration with sub-percent uncertainty with \planck\ data and 2\% uncertainty with \sptpol\ data.
We arrive at similar conclusions when extending $\Lambda$CDM to include the amplitude of lensing $\Alens$, the number of relativistic species $\nnu$, or the sum of the neutrino masses $\mnu$. 
The uncertainties on cosmological parameters are minimally impacted when marginalizing over polarization calibration, except, as can be expected, for the uncertainty on the amplitude of the primordial scalar power spectrum $\lnAs$, which increases by $20-50\%$. 
However, this information can be fully recovered by adding \TT\ data. 
For current and future ground-based experiments, SPT-3G and CMB-S4, 
we forecast the cosmological parameter uncertainties to be minimally degraded when marginalizing over polarization calibration parameters.  
In addition, CMB-S4 could constrain its polarization calibration at the level of $\sim$\,0.2\% by combining TE and EE, and reach $\sim$\,0.06\% by also including TT.
We therefore conclude that relying on calibrating against \planck\ polarization maps, whose statistical uncertainty is limited to $\sim0.5\%$, would be insufficient for upcoming experiments.

\end{abstract}
\maketitle

\section{Introduction}

The $\Lambda$ cold dark matter ($\Lambda$CDM) model has emerged to be the leading model in describing our universe
since the advent of precision measurements of the anisotropies in the cosmic microwave background (CMB).
On the largest angular scales, we have satellite measurements from \wmap\ and \planck\ that reach 
cosmic-variance limits in the temperature anisotropy spectrum up to multipoles $\ell \sim 500$ and $\ell \sim 1600$
respectively~\cite{wmap09,planck2016features, planck2018I}.
On small angular scales, large aperture ground-based experiments like the Atacama Cosmology Telescope (\act) and the South Pole Telescope (\spt) provide high signal-to-noise
measurements of the CMB damping tail~\cite{louis16, story12}, in both temperature and polarization.

As elucidated and forecasted in~\cite{galli2014} and demonstrated by \planck\ and recent results from ground-based telescopes, polarization measurements of the CMB are increasingly dominating over the temperature measurements in terms of their statistical constraining power on cosmological parameters.
However, in order to fully take advantage of these upcoming data sets, systematic errors that could bias the polarization measurements must be sufficiently mitigated and controlled. 
Specifically, recent \planck\ results show that cosmological parameters can be biased by one of the main polarization systematics---errors in the estimates of the polarization efficiencies of the detectors~\cite{planck2018likelihood,planck2018cosmoparam}.
For \planck, the polarization efficiencies of its detectors as measured in-flight were 
discrepant from what were expected from laboratory measurements by up to 5 times the statistical uncertainties of the laboratory measurements~\cite{planck2018III}. 
To account for this discrepancy, the \planck\ polarization calibrations at different frequencies were then re-evaluated by requiring the polarization spectra to recover the $\Lambda$CDM cosmology inferred by the temperature spectrum measurements, effectively modeling the detector polarization efficiencies as overall calibration of the polarization maps per frequency, \pcal.

In this work, we propose an alternative method to extract polarization calibration as a potential cross-check for direct approaches.
Typically, polarization calibration parameters are included in cosmological parameter estimation as
nuisance parameters with priors informed by external calibration steps~\cite[e.g.][]{bicep1calib, actcalib, polocalc}.
Here, we jointly fit the $\Lambda$CDM and extension models to the CMB \TE\ and \EE\ spectra
allowing the polarization calibration parameters to float, i.e., we let the data to self-calibrate \pcal\ given a model. 
We show that the combination of just \TE\ and \EE\ is sufficient in providing a tight \pcal\ constraint, 
and the \pcal\ uncertainty can be further improved by including the temperature power spectrum \TT. 
Atmospheric noise degrades the ground-based \TT\ measurement more than 
satellite \TT\ or ground-based \TE\ and \EE\ measurements. 
For this reason, the ability to self-calibrate \pcal\ with only \TE\ and \EE\ as demonstrated by this work is of particular interest to current and upcoming ground-based experiments~\cite[e.g.][]{bender2018, pb2, advact, so, cmbs4dsr}.

The inferred polarization calibration from our proposed method can produce tight constraints because of the different dependence on \pcal\ of TE and EE, which
breaks parameter degeneracies with other cosmological parameters.
While this inferred polarization calibration is admittedly model-dependent, it is nevertheless useful as a consistency check against polarization calibration estimated through other methods.
Furthermore, we show that most $\Lambda$CDM parameter constraints are only mildly to negligibly degraded
when marginalizing over \pcal, and common extensions to $\Lambda$CDM are insensitive to marginalizing over this
extra parameter.

In the following, we apply this method to \sptpol\ and \planck\ data and show that \pcal\ are constrained
to percent level precision for these experiments across $\Lambda$CDM and its extensions, including
the lensing amplitude $\Alens$, the effective number of relativistic species $\nnu$, and 
the sum of neutrino masses $\mnu$. 
We take inputs from a recent \sptpol\ power spectrum analysis~\cite[][hereafter H18]{henning18} and 
\planck's latest data release~\cite{planck18overview} and sample parameter spaces without imposing priors on their respective polarization calibration parameters. 
With the recent release of the ACTpol DR4 data, we apply this method to the publicly available {\tt ACTPollite} likelihood~\cite{choi20,aiola20}
to demonstrate the ease of application of this approach.
We use {\sc CosmoMC}~\cite{cosmomc} for sampling the posterior distributions of \sptpol\ and \planck, and {\sc Cobaya}~\cite{cobaya} for ACTpol.
To check the relevance of this method for upcoming and future data sets, 
we forecast the \pcal\ uncertainty and the changes in cosmological parameter uncertainties when marginalizing over \pcal\ 
for SPT-3G and CMB-S4.
While this paper was in its final stages of preparation, the results from the first season of the SPT-3G experiment were released \cite{eete2018}. We leave the application of our method to this data set to future work.

This paper is organized as follows. 
In \refsec{PE}, we summarize polarization calibration as defined in \sptpol\ and \planck. 
We present results for \sptpol, ACTpol, and \planck\ in Sections~\ref{sec:sptpol} to~\ref{sec:planck}. 
Our forecasts for SPT-3G and CMB-S4 are detailed in~\refsec{forecasts}.
We conclude in~\refsec{conclusions}.

\section{Polarization efficiency and effective calibration}
\label{sec:PE}

The power absorbed by a polarized detector in an experiment such as \planck\ or  \sptpol\ at time $t$ can be modeled as:
\begin{equation}
P(t) \hspace{-0.8mm} = \hspace{-0.8mm} G\left\{I + \rho\left[ Q\cos 2(\psi(t))+U \sin 2(\psi(t) ) \right] \right\} + n(t), \ \ \label{bol}
\end{equation}
where $I$, $Q$, and $U$ are the Stokes parameters that characterize the intensity 
and polarization fields,  $G$ is the effective gain (setting the
absolute calibration), $\rho$ is the detector polarization efficiency,
$\psi(t)$ is the angle of the detector with respect to the sky and $n(t)$ is the detector noise. 
Here we have omitted effects from beams and bandpasses without loss of generality.

Intensity and polarization $I$, $Q$, and $U$ maps per frequency are then produced via map-making \cite[e.g.,][]{planck2018III} by co-adding observations at different times and from different detectors. 
Relative calibration corrections are applied across detectors and the co-addition is weighted given the noise of the time-ordered data over some observing period. 
In the following, we focus on the impact of errors in the estimate of detector polarization efficiency at the coadded map level, which 
can be effectively captured at each frequency by a polarization calibration correction parameter \pcal.\footnote{The 
polarization calibration correction parameter, \pcal, are sometimes called polarization efficiency corrections in \planck\ papers. 
Unless specifically referring to detector polarization efficiencies, we use polarization calibration \pcal\
as applied at the map level to refer to this correction. 
In this paper, we would often shorten ``polarization calibration correction parameter" to polarization calibration.}

For the \sptpol\ TE and EE analysis in H18, 
polarization maps are first made incorporating detector polarization efficiencies and angles measured on ground.
Then, before forming data power spectra, the temperature and polarization maps are calibrated against \planck\ maps.
The calibration factors $\epsilon$ are formed by first taking the ratio of the cross-spectrum between two halves of \sptpol\ maps and
the cross-spectrum between \planck\ maps and \sptpol\ maps.  
The \planck\ maps are masked and filtered identically as the \sptpol\ maps and thus have the same
filter transfer function and mode-coupling. 
The remaining differences from the beams $B_{b}$ and the pixel-window function $\sqrt{F_{b}}$
of the input \planck\ maps are accounted for as follows:
\begin{equation}
\label{eqn:sptcal}
\epsilon_{b} = \frac{\sqrt{F^{Planck}_{b}} B^{Planck}_{b}}{B^{\rm SPT}_{b}} \frac{C_{b}^{{\rm SPT}_i \times {\rm SPT}_j}}{C_{b}^{{\rm SPT} \times Planck}},
\end{equation}
where subscript $b$ denotes binned multipole, and $i,j$ denote different halves of the \sptpol\ data.
The calibration factors are extracted by averaging across the multipole ranges $600 < \ell < 1000$ for temperature and $500 < \ell < 1500$ for polarization.
The \planck\ DR2 \commander\ polarization maps are used to obtain the polarization calibration factor, and provide a $\sim$6\% correction to the $Q$ and $U$ maps (see sections 4.5.2 and 7.3 in H18 for further details).   
The uncertainties of the calibration factors are incorporated when sampling cosmological and nuisance parameters.
Specifically, the theoretical spectra to which the data are compared are scaled by $1/$(\tcal$^2$ \pcal ) for \TE\ and $1/$(\tcal \pcal)$^2$ for \EE, 
where \tcal\ denotes the overall residual calibration of the maps and \pcal\ denotes the polarization calibration correction. 
Gaussian priors with mean of unity and uncertainties of 0.34\% and 1\% are applied 
to \tcal\ and \pcal\ respectively, based on the uncertainties of the ratio estimates in~\refeqn{sptcal}. 
It is the prior on \pcal\ that we remove in this work.

For \planck, the modeling of polarization calibration is different from the one used in H18 in two ways. 
First, the \planck\ likelihood at high-$\ell$\footnote{The high-$\ell$ likelihood covers $\ell>30$. We assume here that polarization efficiency corrections have a negligible impact on the low-$\ell$ polarization likelihood due to the large uncertainties in this regime due a combination of cosmic variance, noise, and systematic uncertainties.} includes maps from 3 frequencies, 100, 143, and 217\,GHz, in constrast to the single-frequency analysis done in H18 at 150\,GHz. 
Second, 
while the \sptpol\ \pcal\ is defined at the map level, the  \planck\ effective polarization calibration parameters $\calibC^{EE}_{\nu}$ are defined at the power spectrum level for each frequency spectrum $\nu\times\nu$ used in the high-$\ell$ likelihood.\footnote{Thus, the polarization efficiency for a cross-frequency spectrum $\nu\times\nu'$ in, e.g., $EE$ is $\sqrt{\calibC^{EE}_{\nu}\times \calibC^{EE}_{\nu'}}$.} Thus, \pcal $ \, =\sqrt{\calibC^{EE}}$ for each frequency.

Specifically, the theory power spectra to which the data is compared are multiplied by a calibration factor $g$ defined as
\begin{eqnarray}
g_{\nu\times\nu'}^{XY} &=& \frac{1}{2\calibM_{\rm P}^2}\left( \frac{1}{\sqrt{\calibC^{XX}_{\nu}\calibC^{YY}_{\nu'}}} + \frac{1}{\sqrt{\calibC^{XX}_{\nu'}\calibC^{YY}_{\nu}}} \right) \,. \label{eq:caldef}
\end{eqnarray}
Here, $\nu\times\nu'$ indicate the frequency spectra with $\nu,\nu'=100, 143, 217$\,GHz; the spectra are then either for $XY=TE$ or $XY=EE$.  
$\calibC^{TT}_{\nu}$ denotes temperature calibration parameters, which are separately determined and on which priors are set. 
$\calibC^{TT}_{143}$ is set to unity so that the 143\,GHz temperature map is taken as a reference. 
Finally, $\calibM_{\rm P}$ is the overall \planck\ calibration parameter defined at the map level, on which a Gaussian prior\footnote{We denote Gaussian priors with mean $\mu$ and standard deviation $\sigma$ as $(\mu,\sigma^2)$, and uniform priors between $v_\mathrm{min}$ and $v_\mathrm{max}$ as $[v_\mathrm{min},v_\mathrm{max}]$.} of  $\calibM_{\rm P} = (1,0.0025^2)$ is set (see Section~3.3.4 of ~\cite{planck2018likelihood} for further details). 
As detailed in~\refsec{planck}, in the baseline \planck\ analysis, $\calibC^{EE}_{\nu}$ are fixed to the values obtained by comparing the EE data spectra to the theory spectra computed given the best-fit cosmology to the TT spectra.
In this work, $\calibC^{EE}_{\nu}$ are nuisance parameters to be constrained by the data themselves.
Given the different definitions of the polarization calibration in these \sptpol\ and \planck\ works, in the rest of this paper we will always specify whether the quoted uncertainties refer to the map-level (\pcal) or power-spectrum level ($\calibC^{EE}_{\nu}$) corrections. 
In~\refsec{planck}, we will provide results for the \planck\ data using both definitions.

\section{\sptpol}
\label{sec:sptpol}

\subsection{Data and model description}

We use the \sptpol\ \TE\ and \EE\ power spectrum measurements from H18.
The generation of these measurements is described in detail in H18 and here we highlight relevant aspects of that work. 
Data in H18 came from the 150\,GHz band observations made by the \sptpol\ camera on the South Pole Telescope
over an effective area of 490 deg$^2$.
The power spectra cover angular multipoles $\ell$ between 50 and 8000.
The polarization noise level measured in the range $1000 < \ell < 3000$ of this data set is $9.4\,\mu$K\,arcmin.

For the $\Lambda$CDM baseline case, we sample the identical model space as in H18 
using the same  covariance matrix with {\sc CosmoMC}~\cite{cosmomc}. 
The model parameter space is composed of $\Lambda$CDM, foreground, and nuisance parameters.
The $\Lambda$CDM parameters are 
the cold dark matter density $\Omega_ch^2$; the baryon density $\Omega_bh^2$;
the amplitude and tilt of the primordial scalar power spectrum $\lnAs$ and $n_s$;
the optical depth to reionization $\tau$;
\textsc{CosmoMC}'s internal proxy to the angular scale of the sound horizon at decoupling, $\theta_{MC}$.
A Gaussian prior is set on $\tau: (0.0544, 0.0073^2)$ given the \planck\ results~\cite{planck18cosmoparam}. 
The sum of neutrino mass $\mnu$, when not sampled, is fixed to 0.06 eV. 
On the other cosmological parameters, we set large uniform priors.

We consider Galactic dust foregrounds and the extragalactic foregrounds from polarized point sources.
We model and set priors for them identically as in H18.
The priors on the amplitudes of dust at $\ell$ of 80, $A^{TE}_{80}$ and $A^{EE}_{80}$, are set to be uniform with $[0,2\mu$K$^2]$;
the priors on the spatial spectral indices, $\alpha_{TE}$ and $\alpha_{EE}$, are set to $(-2.42, 0.02^2)$.
Finally, the prior on the amplitude of polarized sources $D^{\rm PS_{EE}}_{3000}$ is set to $[0,2.5 \mu$K$^2]$. 

As in H18, the nuisance parameters are beam uncertainties, super-sample lensing~\cite{2014PhRvD..90b3003M}, and temperature and polarization calibrations. 
We include effects from super-sample lensing with the prior on $\kappa$ to be $(0.0, 0.001^2)$. 
We model beam uncertainties using two eigenmodes with prior $(0.0, 1^2)$ on each mode. 
The overall residual calibration parameter $T_{\rm cal}$ has prior $(1.0, 0.0034^2)$.
Finally, as for the focus of this paper $P_{\rm cal}$, we either set a prior of $(1.0, 0.01^2)$, which
is the baseline of H18, or no prior, which is the method we propose to
let \pcal\ be determined by the data.

In the following, we will report results obtained either from \TE\ and \EE\ separately, or from the combination of the two, which we will refer to as \TE,\EE.

\subsection{Main results}

\begin{figure}
\begin{center}
\includegraphics[width=0.48\textwidth]{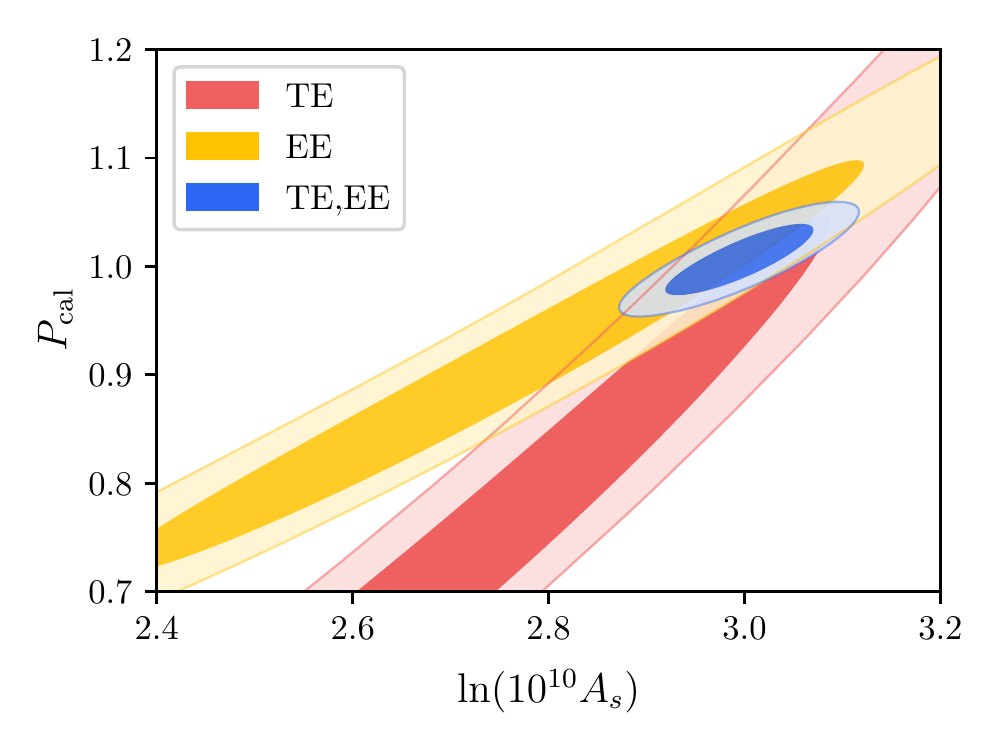}
\caption{ln($10^{10} A_s$) vs \pcal\ in $\Lambda$CDM for \sptpol\ \TE, \EE, and \TE,\EE, with no \pcal\ priors. 
We exploit the different degeneracy directions between $\lnAs$ and \pcal\ from \TE\ and \EE\ to constrain \pcal.  
}
\label{fig:spt_pcal_vs_as}
\end{center}
\end{figure}

To illustrate the idea,
in~\reffig{spt_pcal_vs_as}, we show the 2D posterior of $\lnAs$ and \pcal\
from TE, EE, and TE,EE without imposing a \pcal\ prior.
We see that without a \pcal\ prior, the constraints on $A_s$ from TE alone and EE alone
are very degenerate
with \pcal.
However, since the \pcal\ dependence from TE and EE are different (linear versus quadratic in \pcal\ 
respectively), the combined TE,EE constraint on $A_s$ and \pcal\ without a prior
are significantly reduced. 
This illustrates the potential of combining the TE and EE spectra in constraining
\pcal\ without significantly degrading constraints on $\Lambda$CDM parameters.
Furthermore, we find that the \pcal\ parameter as sampled is consistent with unity. 
This serves as cross-check to the polarization calibration determined by the comparison to the \planck\  \commander\ polarization maps.
In the following, we first show that the constraints on \pcal\ are sufficiently precise
and stable across different models to be used as a cross-check for other sources of measurements.
We then discuss effects on cosmological parameter uncertainties when marginalizing over \pcal. 

\begin{figure}
\begin{center}
\includegraphics[width=0.4\textwidth]{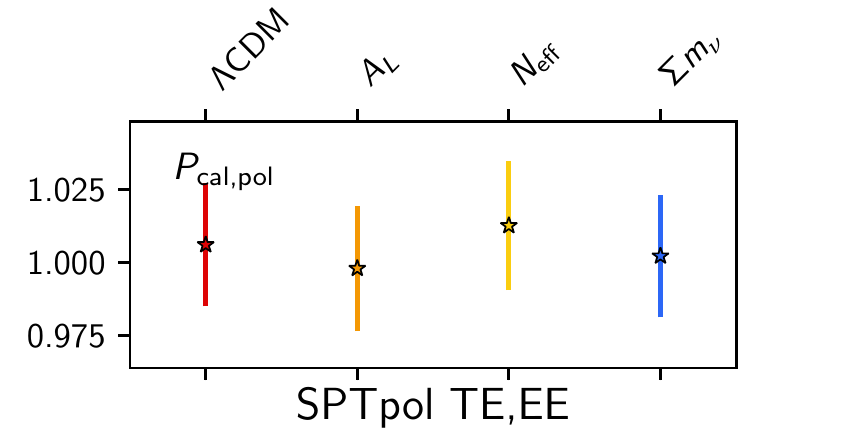}
\caption{Marginal mean and 68\% confidence level error bars 
on \pcal\ 
obtained from \sptpol\ \TE,\EE\ data assuming the $\Lambda$CDM model and a few of its extensions. 
The determination of \pcal\ is only slightly affected by the choice of cosmological model. }
\label{fig:sptpolPE}
\end{center}
\end{figure}

For this \sptpol\ data set, we obtain a $\sim2\%$ constraint on \pcal\ in $\Lambda$CDM and 
three extensions---$\Alens$, $\nnu$, and $\mnu$, as listed in~\reftab{sptpolPE} and shown
in~\reffig{sptpolPE}.
This level of precision is sufficient to cross-check the baseline approach used in H18
in which the \sptpol\ polarization maps are calibrated against the \planck\ \commander\ maps.
In other words, without applying the polarization calibration correction from comparing against \planck,
one would arrive at a similar conclusion that a 6\% correction should be applied to the calibration of the
polarization maps if one lets \pcal\ float while sampling the $\Lambda$CDM and extension model spaces 
with the TE,EE data set.
We note that in all three extension scenarios, the \pcal\ constraint does not degrade significantly,
which shows that this approach is useful as cross-checks beyond just the $\Lambda$CDM model. 

\begin{table}[t]
\centering
\caption{Polarization calibration parameters obtained from \sptpol\ 
data assuming different  models.
For reference, using the baseline \pcal\ prior in H18 of 1\%, we find \pcal\ = 1.0015$\pm$0.0090
for the \lcdm\ model.} 
\label{tab:sptpolPE}
\begin{tabular}{c|c }
Model           & \sptpol\ TE,EE   (no \pcal\ prior)      \\ 
\hline
\hline
$\Lambda$CDM &  	1.0061 $\pm$ 0.0210 \\
$\Lambda$CDM+$\Alens$ & 0.9980 $\pm$ 0.0216 \\
$\Lambda$CDM+$\nnu$ & 1.0126 $\pm$ 0.0222 \\
$\Lambda$CDM+$\mnu$ & 1.0022 $\pm$ 0.0209  \\
\end{tabular}%
\end{table}

\begin{figure}
\begin{center}
\includegraphics[width=0.5\textwidth]{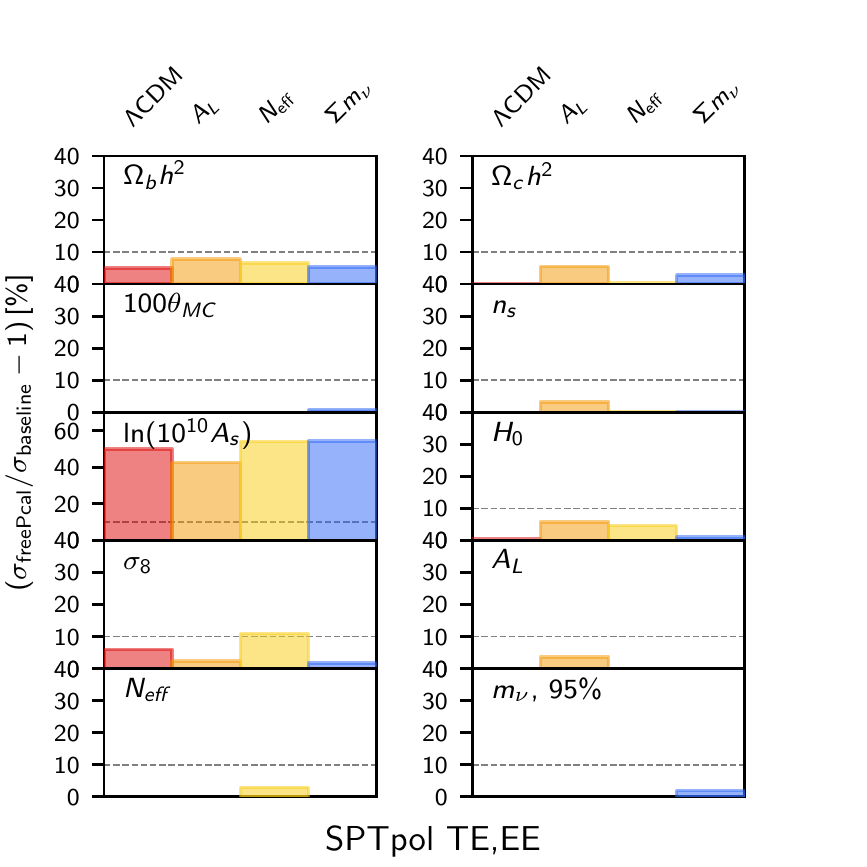}
\caption{Impact of freeing \pcal\ on the error bars of cosmological parameters for the \sptpol\ \TE,\EE\ data . We show the ratio of the error bars obtained letting the \pcal\ parameter free to vary, over the ones obtained using the baseline \sptpol\ settings, in units of percent, $\sigma_\mathrm{free \,Pcal}/\sigma_\mathrm{baseline}-1\,[\%]$. The horizontal dashed line indicates a 10\% increase in the error bars. We show results for the $\Lambda$CDM model and a few of its extensions. Only the constraints on $\lnAs$ are significantly weakened by letting \pcal\ free to vary. }
\label{fig:sptpol_cosmo}
\end{center}
\end{figure}

The stable uncertainties on \pcal\ across $\Lambda$CDM and the few extensions suggest that 
\pcal\ has little degeneracy with other parameters. 
Indeed, most cosmological parameter constraints are only negligibly to mildly degraded
when we relax the \pcal\ prior for the \sptpol\ TE,EE data set.
We show in~\reffig{sptpol_cosmo} the ratios of cosmological parameter uncertainties between the no \pcal\ prior
and the baseline \pcal\ prior case for the models considered.
The constraints on $A_s$ degrade most, by $40-60\%$ depending on the model.
This is expected given the correlation between $\lnAs$ and \pcal.
The correlation\footnote{We define the correlation between two parameters $x,y$ as $\rho_{x,y}=\mathrm{cov}(x,y)/\sqrt{\mathrm{cov}(x,x)\mathrm{cov}(y,y)}$, with $\mathrm{cov}(x,y)$ the elements of the parameter covariance matrix.} is 84\% for the $\Lambda$CDM case, as suggested in~\reffig{spt_pcal_vs_as}. 
All of the rest of the parameter uncertainties increase by $\lesssim 10\%$ when marginalizing over the broadened
\pcal\ posterior space. 
We show in~\refsec{forecasts} that the degradation in $A_s$ disappears if we include the temperature
spectrum measurement \TT\ as part of the input.
This is because \TT\ tightly constrains $A_s$ independent of \pcal. 
For data sets similar to \sptpol, not only are the constraints on \pcal\ precise enough for cross-checks with
other approaches, most cosmological parameter constraints are also minimally degraded when no \pcal\ priors are imposed.

As one way of demonstrating consistency, we compare the inferred \pcal\ values from the
TE-only and EE-only data sets when the rest of the parameters are fixed to the best-fit from the TE,EE joint fit
in $\Lambda$CDM with the baseline \pcal\ prior. 
The marginalized \pcal\ are 
\pcal =$0.997 \pm 0.020$ and
\pcal =$0.991 \pm 0.005$ 
for the TE and the EE data set respectively.
This shows that the individual data set does not prefer a statistically different
\pcal; there is no significant systematic residuals that project onto \pcal.

\subsection{The $\Alens$ case}

\begin{figure}
\begin{center}
\includegraphics[width=0.5\textwidth]{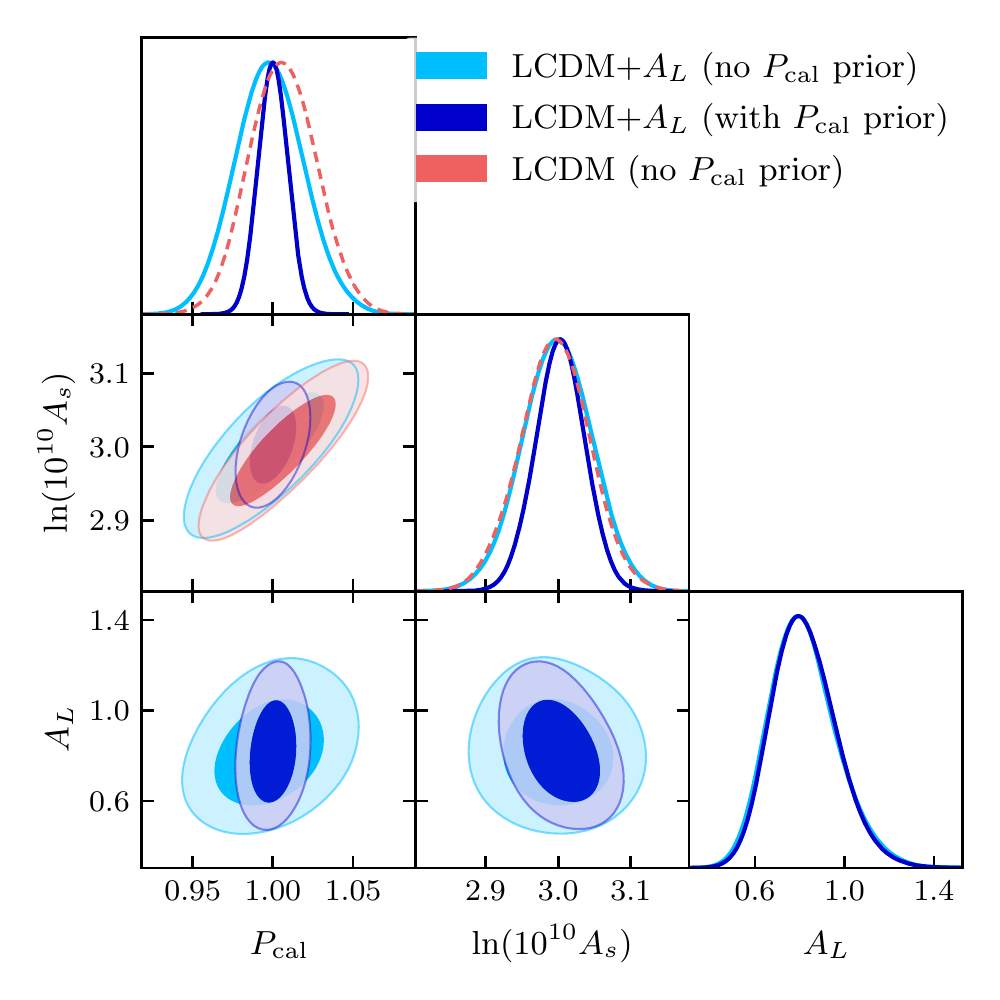}
\caption{$A_s$, \pcal, and $\Alens$ posteriors with and without 
$\Alens$ free for the \sptpol\ \TE,\EE\ data set.
The uncertainty on $\Alens$ is unchanged with and without the \pcal\ prior,
ensuring the strong $\Alens$ constraint from polarization-only spectra
even when freeing \pcal. 
}
\label{fig:spt_alens_1d}
\end{center}
\end{figure}

We now turn to one particularly interesting parameter extension, $\Alens$, a non-physical parameter
that tunes the effect of gravitational lensing on the CMB primary spectra, changing the amount of smoothing of its peaks and troughs~\cite{calabrese2008}.
In \planck, an excess peak smoothing was observed in their temperature power spectrum
at the 2.8\,$\sigma$ level compared with the $\Lambda$CDM expectation~\cite{planck18cosmoparam}.
One key way of differentiating whether the excess smoothing is a statistical fluctuation, the result of unmodeled systematic errors, or new physics
is to test if this trend persists in polarization~\cite[e.g.][]{aiola20}.
One concern of our method would be that marginalizing over the no-prior \pcal\ would degrade the $\Alens$ 
constraint enough that one can no longer tell if the polarization data show similar trends.
As shown in~\reffig{spt_alens_1d}, the constraints on $\Alens$ for TE,EE with and without \pcal\ priors 
are almost identical, retiring related concerns. 

As an aside, we note that the lensing information from peak smoothing reduces what would otherwise be almost complete degeneracy between \pcal\ and $A_s$ in the TE-only and EE-only cases.
Specifically, because peak smoothing provides a second handle for measuring $A_s$, the 
$\Lambda$CDM TE-only and EE-only constraints on \pcal\ are 21\% and 13\% respectively 
(as shown in the red and yellow contours in \reffig{spt_pcal_vs_as}).
 By contrast, when the peak-smoothing information is absorbed by the additional parameter $\Alens$, 
 the \pcal\ constraints degrade by more than a factor of two in both \TE-only and \EE-only cases.

\section{ACTpol}

We apply this method to the recent ACTpol DR4 data set
on the frequency-combined CMB-only spectra, using the {\tt ACTPollite} likelihood~\cite{choi20, aiola20}.
We note that the flat prior applied on $y^P$, the ACTpol polarization calibration parameter,  
is sufficiently broad ([0.9, 1.1])
that it is already allowing $y^P$ to float to that extent. 
Here we estimate how well this ACTpol data set can constrain polarization calibration using
just the \TE\ and \EE\ spectra, with the prior on $y^P$ further widened.
We also check if the \TE,\EE\ $y^P$ result is consistent with the \TT,\TE,\EE\ $y^P$ result.

We use only the \TE\ and \EE\ frequency-combined spectra
without \TT\ on both the wide and the deep patch. 
We then transform the $y^P$ samples by applying an inverse to match
the \pcal\ definition, \pcal$=1/y^P$.
With this setup, we find \pcal\ $= 1.0113 \pm 0.0150$ in the $\Lambda$CDM model.
It is consistent with the $y^P$ result from~\cite{aiola20}, which includes the TT spectra, of $y^P = 1.0008 \pm 0.0047$.

\section{\planck}
\label{sec:planck}
\subsection{Data and model description}
\label{sec:planckdata}
In this section, we test whether jointly fitting the \planck\ TE and EE spectra with no prior on \pcal\ would produce sufficiently precise \pcal\ measurements to serve as useful cross-checks for other approaches. We also test the level of impact of this approach on the uncertainties on cosmological parameters.

In \planck, polarization efficiencies, as well as polarization angles, were measured on the ground in \cite{rosset2010} and taken into account in the map-making algorithm \sroll~\cite{planck2018III}. At the frequencies used in the high-multipole likelihood (100, 143, 217\,GHz), polarization efficiencies per detector were found to be between $83\%$ and $96\%$, with estimated uncertainties between $0.1$ and $0.3\%$ at the map level. 
However, tests performed on the maps, which compared strongly emitting polarized galactic dust regions as observed by different detectors, suggested that residual polarization efficiency errors are several times larger than the expected uncertainties reported in~\cite{rosset2010}, as shown in \cite{planck2018III}. 
Left uncorrected, these residuals in the polarization efficiencies can impact cosmological parameters up to fractions of a sigma by biasing the overall amplitude of the $\TE$ and $\EE$ spectra used in the high-multipole likelihood.

In order to correct for this effect,  effective polarization calibrations were estimated by the \planck\ collaboration by comparing the \TE\ and \EE\ power spectra at 100, 143, and 217\,GHz to fiducial \TE\ and \EE\ spectra computed from the \LCDM\ best-fit to the \TT\ data. 
Polarized galactic contamination was cleaned using information from the 353\,GHz channel~\cite{planck2018likelihood}. The fits were performed on a limited range of multipoles ($\ell=200-1000$) to discard regions affected by foreground cleaning or noise uncertainties 
and over about $\sim 60\%$ of the sky (see \cite{planck2018likelihood} for details). 
The advantage of this method is that it provides an absolute reference with small uncertainties. 
The disadvantage is that the polarization efficiency corrections found in this way depend on the cosmological model fitted to the temperature data (although this was tested to have a small impact). 
This method enabled determinations of the polarization calibration for \EE\  with uncertainties below  $\lesssim 0.5\% $ at the map level ($\lesssim 1\% $ at power spectrum level) and for \TE\ with uncertainties below $\lesssim 1\% $ ($\lesssim 2\% $) in each of the three frequencies used in the high-multipole \planck\ likelihood. 
Up to a global polarization calibration, the derived $\calibC^{EE}_{\nu}$s were found to be consistent with the results of the component separation algorithm SMICA~\cite{planck2018III}, which measures relative inter-frequency calibration ratios between foreground-cleaned polarization maps.
Furthermore, in~\cite{planck2018likelihood}, it was noted that the estimates obtained separately from \EE\ and \TE\ should agree given the same polarization maps.
However, the two measurements were found to differ by up to $1.7\pm1\%$ at the map level at 143\,GHz (see Section~3.3.4 of \cite{planck2018likelihood}).
As we will show below, this difference cannot be reconciled by the approach we propose in this work---leaving polarization efficiencies to freely vary.
Since the difference in polarization calibration from \TE\ and \EE\ is small enough that it could be caused by statistical fluctuations,
we leave the investigation of potential biases to parameters to future work and 
focus on the constraints on \pcal\ given the \planck\ data set and impact on cosmological parameters.
 
We consider the 2018 final release of the \planck\ data~\cite{planck2018likelihood}. 
We use the  low-multipole likelihood in polarization \simall\ ($\ell=2-29$ in \EE\ only), which we will refer to as ``\lowE." 
For high multipoles, we use the \plik\ likelihood ($\ell=30-1997$ in \EE\ and \TE), which we will refer to as \TE\ and \EE\ separately or \TE,\EE\ when used in combination. 
For cross-checks, we use the \TT\ \commander\  likelihood at low-$\ell$ ($\ell=2-29$) and \plik\ at high-$\ell$ ($\ell=30-2508$) and we refer to the combination of the two as \TT.
We model polarization calibration only for the high-$\ell$ likelihoods, 
because their impact on low-$\ell$ spectra are negligible compared to cosmic variance, noise, and systematic uncertainties in this multipole range.
In the baseline \planck\ results using the \plik\ likelihood, 
the polarization calibration to the \TE\ and \EE\ spectra are fixed
to the ones obtained from comparing the \EE\ spectra at different frequencies to the $\Lambda$CDM best-fit of the \TT+\lowE\ data combination.
These baseline parameters are listed in Table~\ref{tab:planckPE}. 

\subsection{Main results and robustness assessment}

\begin{figure}
\begin{center}
\includegraphics[width=0.48\textwidth]{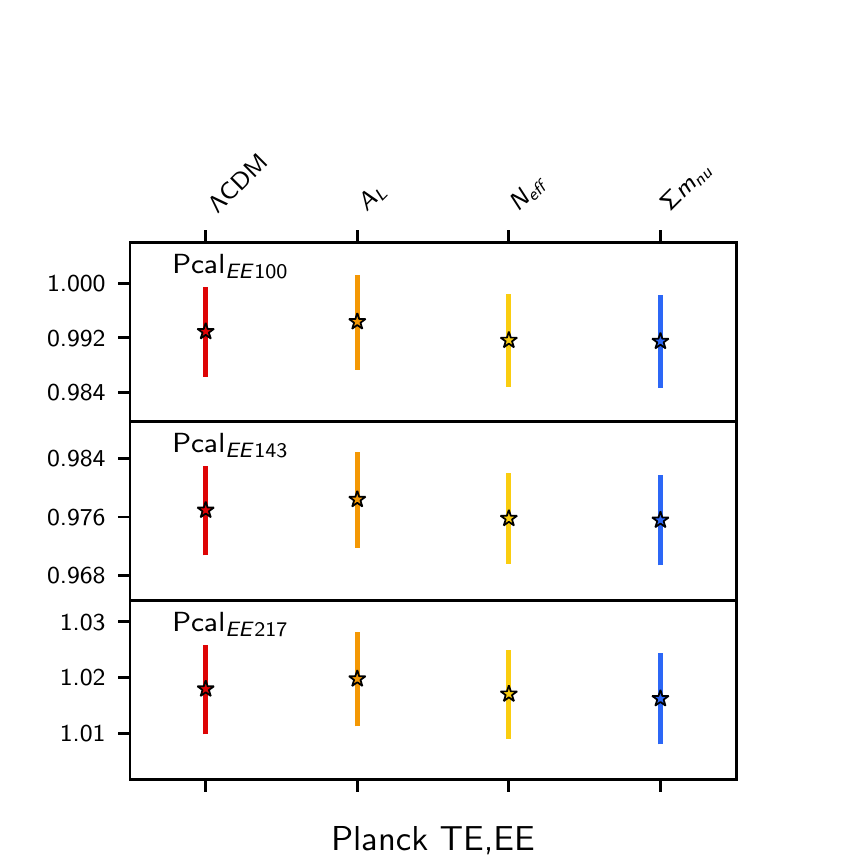}
\includegraphics[width=0.48\textwidth]{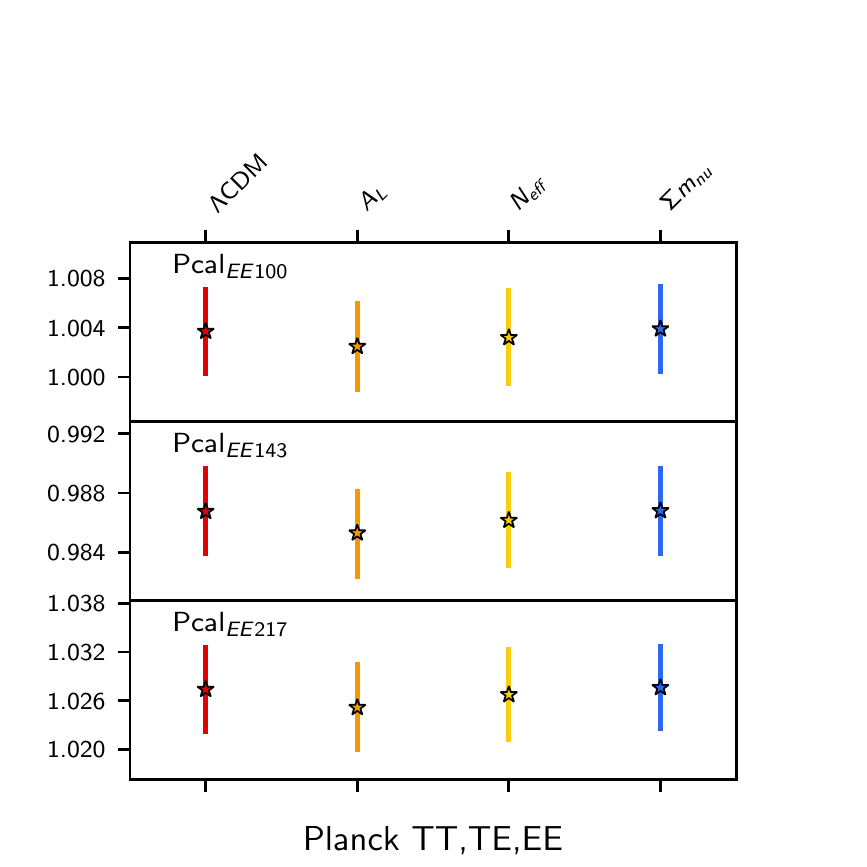}

\caption{Marginal mean and 68\% confidence level error bars on the three \planck\ \pcal\ frequency parameters when they are let free to vary assuming different cosmological models. The top plot shows the results for \planck\ \TE,EE, while the bottom one shows \planck\ \TT,TE,EE. Estimates on the \pcal\ parameters do not change significantly when varying the cosmological model.}
\label{fig:planckpolTTTEEE}
\end{center}
\end{figure}

We first discuss the uncertainties on \pcal\ for the \planck\ data set when it is free to vary.
Using  \TE,\EE+\lowE, we find one can determine the polarization calibrations with uncertainties smaller than $\sim1\%$ at the map level. 
More specifically we find uncertainties of $0.65\%$, $0.6\%$ and $0.8\%$ at the map level for $\nu = 100, 143, 217$\,GHz  respectively (corresponding to $1.3\%$, $1.2\%$ and $1.7\%$ at the power spectrum level).
Furthermore, we compare these uncertainties to the ones obtained with the \TT\ power spectra included. 
We find that the error bars shrink by almost a factor of $2$ to $0.35\%$, $0.31\%$ and $0.51\%$ at the map level for the three frequencies and similarly at the power-spectrum level. 
The measurements and uncertainties are reported in~\reftab{planckPE} and shown in~\reffig{planckpolTTTEEE}.
With and without TT, the uncertainties on the \pcal\ factors are comparable to ones used in the \plik\ likelihood. 
This demonstrates that this approach yields relevant constraints on \pcal\ for cross-checks of other approaches.

\begin{table}[t]
\centering
\caption{Polarization calibrations at power spectrum level obtained from \planck\ data assuming different  cosmological models. We also report the corresponding polarization calibrations at map level (\pcal$=\sqrt{\calibC^{EE}}$, $\sigma($\pcal$)\sim (\calibC^{EE})^{-1.5}\sigma(\calibC^{EE})/2.$), to ease the comparison with those obtained for \spt\ in Section \ref{sec:sptpol}. The column "baseline" lists the fixed values used in the baseline \planck\ likelihood, which were determined with an uncertainty of $\sim1\%$ at 
the power-spectrum level ($\sim 0.5\%$ at the map level).}
\label{tab:planckPE}
\begin{tabular}{c|c|c|c}
Parameter               & \planckTEEE & \planckTTTEEE   & baseline           \\ 
\hline
\hline
$\Lambda$CDM\\
\hline
$c_{EE100}$ 	& 0.985 $\pm$ 0.013 &  1.007 $\pm$ 0.007 & 1.021\\
$c_{EE143}$	& 0.954 $\pm$ 0.012 & 0.973 $\pm$ 0.006 & 0.966 \\
$c_{EE217}$    & 1.036 $\pm$ 0.017 & 1.056 $\pm$ 0.011 & 1.04\\
\hline
\pcal$^{EE100}$ &0.9925	$\pm$ 0.0066&	1.0035 $\pm$	0.0035\\
\pcal$^{EE143}$ &0.9767	$\pm$ 0.0064&	0.9864 $\pm$	0.0031\\
\pcal$^{EE217}$ &1.0178	$\pm$ 0.0081&	1.0276 $\pm$	0.0051\\

\hline
\hline
$\Lambda$CDM+$\Alens$\\
\hline
$c_{EE100}$ 	& 0.989 $\pm$ 0.014  &1.005 $\pm$  0.0074\\
$c_{EE143}$    & 0.957 $\pm$ 0.013  &0.971 $\pm$  0.0060\\
$c_{EE217}$    & 1.040 $\pm$ 0.017  &1.050 $\pm$  0.012\\
\hline
\pcal$^{EE100}$ &0.9945 $\pm$ 	0.0071	&1.0025 $\pm$	0.0037\\
\pcal$^{EE143}$ &0.9783 $\pm$ 	0.0069	&0.9854	$\pm$ 0.0031\\
\pcal$^{EE217}$ &1.0198 $\pm$ 	0.0080	&1.0247	$\pm$ 0.0056\\
\hline
\hline
$\Lambda$CDM+$\nnu$\\
\hline
$c_{EE100}$ 	& 0.983 $\pm$ 0.013  &1.006 $\pm$  0.0080\\
$c_{EE143}$    & 0.957 $\pm$ 0.012  &0.973 $\pm$  0.0064\\
$c_{EE217}$    & 1.040 $\pm$ 0.016  &1.054 $\pm$  0.012 \\
\hline
\pcal$^{EE100}$ &0.9915 $\pm$	0.0067&	1.0030 $\pm$	0.0040 \\
\pcal$^{EE143}$ &0.9783 $\pm$	0.0064&	0.9864 $\pm$	0.0033 \\
\pcal$^{EE217}$ &1.0198 $\pm$	0.0075&	1.0266 $\pm$	0.0055 \\

\end{tabular}%
\end{table}

In \reftab{planckPE}, we observe shifts in the mean values of the polarization calibrations when \TT\ are added to \TE\ and \EE.
To check that the shifts are consistent with statistical fluctuations, we employ the formalism described in \cite{gratton2019}, which is applicable for comparing two data sets in which one is a subset of the other. 
We find that the observed shifts are consistent with statistical fluctuations at better than the $2\sigma_{\rm exp}$ level, with $\sigma_{\rm exp}=\sqrt{\sigma_{\TE,\EE}^2-\sigma_{\TT,\TE,\EE}^2}$. Finally, we note
that the mean values recovered from the \TT,\TE,\EE\ combination are slightly different from the ones used in the baseline
because of statistical fluctuations due to the different multipole range and sky mask used in the two cases (see also the discussion in section 3.7 of ~\cite{planck2018likelihood}).

We further check how much the constraints degrade when we  exclude the cross-frequency spectra and only use  the combination of the \TE\ and \EE\ frequency auto-spectra  $100\times100$, $143\times143$, and $217\times217$\,GHz. 
We find in this case comparable constraints on polarization  calibrations to our baseline results. 
Furthermore, if we include \TE\ and \EE\ from only one frequency instead of all three as in our previous cases, i.e., we use only the $100\times100$, $143\times143$, or $217\times217$ GHz power spectra, the uncertainties of the polarization calibrations worsen to $1.1\%,$ $0.75\%$ and $2.1\%$ at the map level ($2.1\%,$ $1.5\%$ and $4.1\%$ at the power spectrum level) respectively. 
The large increase in uncertainty for the $217\times217$\,GHz case is because of the more restrictive $\ell$ range of $500-1996$ used at this frequency, which increases the degeneracies between cosmological parameters and polarization calibrations. 
For the other frequencies, the degradation of the constraint is smaller than a factor of $2$.

\begin{figure*}
\begin{center}
\includegraphics[width=0.48\textwidth]{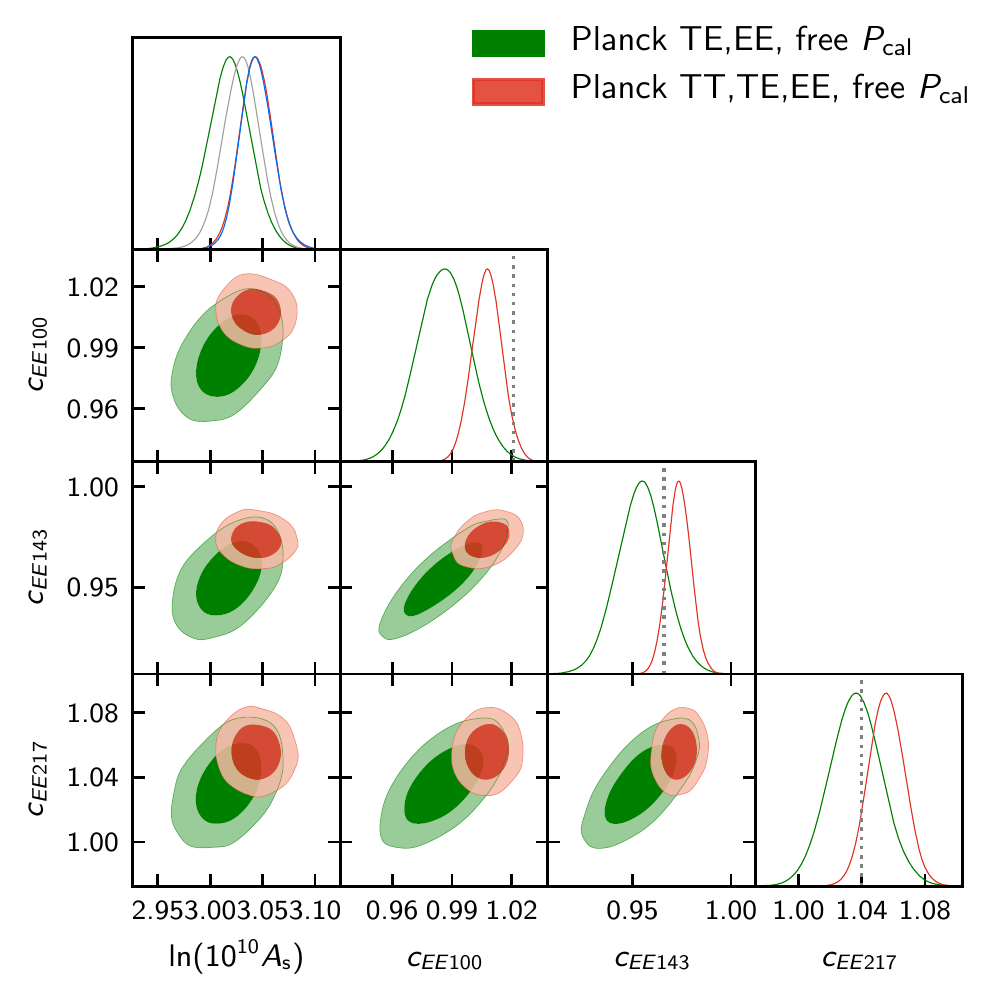}
\includegraphics[width=0.48\textwidth]{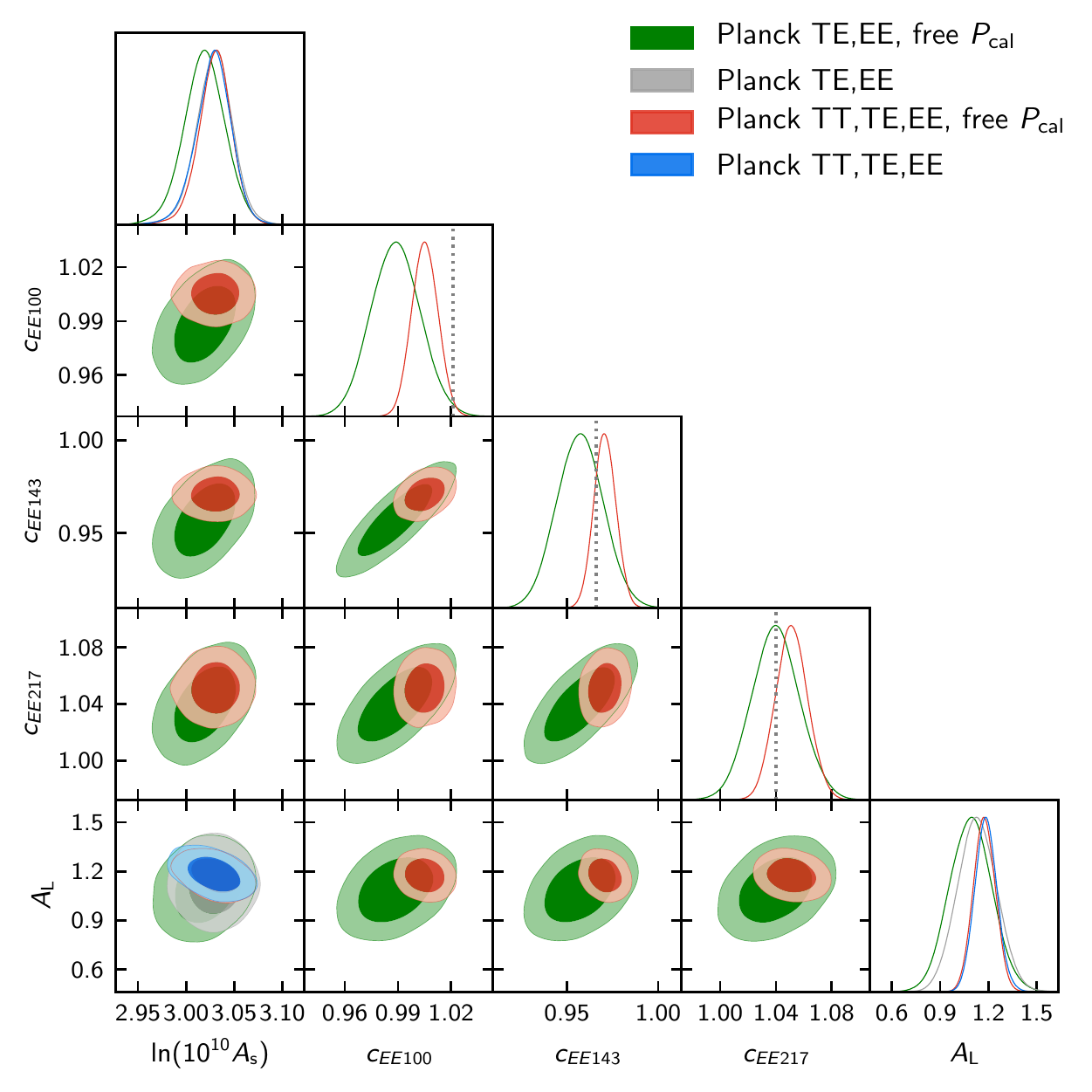}
\caption{One- and two-dimensional posterior distributions of the polarization efficiency parameters and  cosmological parameters for \planck\ \TE,\EE. The left panel shows the results for the $\Lambda$CDM model, while the right panel shows results for the $\Lambda$CDM+$A_L$ model.
}
\label{fig:planck_PE}
\end{center}
\end{figure*}

\reffig{planck_PE} shows the degeneracies between the \pcal\ parameters at different frequencies and the most degenerate  cosmological parameter, $\lnAs$. 
When  using \TE,\EE+\lowE, $\lnAs$ has a $\sim40\%$ correlation with each of the three \pcal\ parameters. 
The second most degenerate parameter is $\Omega_bh^2$ ($\sim30\%$ correlation), while all other parameters have smaller correlations. 
As can be expected, we also find the degeneracies amongst the  \pcal\  parameters to be large: $\rho_{\calibC^{EE}_{100},\calibC^{EE}_{143}}=81\%$, $\rho_{\calibC^{EE}_{100},\calibC^{EE}_{217}}=60\%$ and $\rho_{\calibC^{EE}_{100},\calibC^{EE}_{143}}=66\%$.
These correlations are then lifted when adding information from \TT. 

In terms of the impact on cosmological parameter constraints when allowing \pcal\ parameters to float,
we show the fractional difference in $\Lambda$CDM parameter uncertainties in~\reffig{planckcosmo} for TE,EE 
and~\reffig{planckcosmoTTTEEE} for TT,TE,EE.
Similar to what we see in \sptpol, 
we observe negligible to mild degradation in $\Lambda$CDM parameter uncertainties besides those for $\lnAs$,
given the correlations between the \pcal\ parameters and $\lnAs$.
For the TE,EE data set, the uncertainty of $\lnAs$ increases by $\sim20\%$ when the \pcal\ parameters are 
allowed to float. 
Once TT is included, which independently constrains $\lnAs$, we see that floating \pcal\ has negligible 
impact on all $\Lambda$CDM parameters. 
We will see similar trends in our forecasts in~\refsec{forecasts}.

\subsection{Extended models}
\label{sec:planckext}
\begin{figure}
\begin{center}
\includegraphics[width=0.5\textwidth]{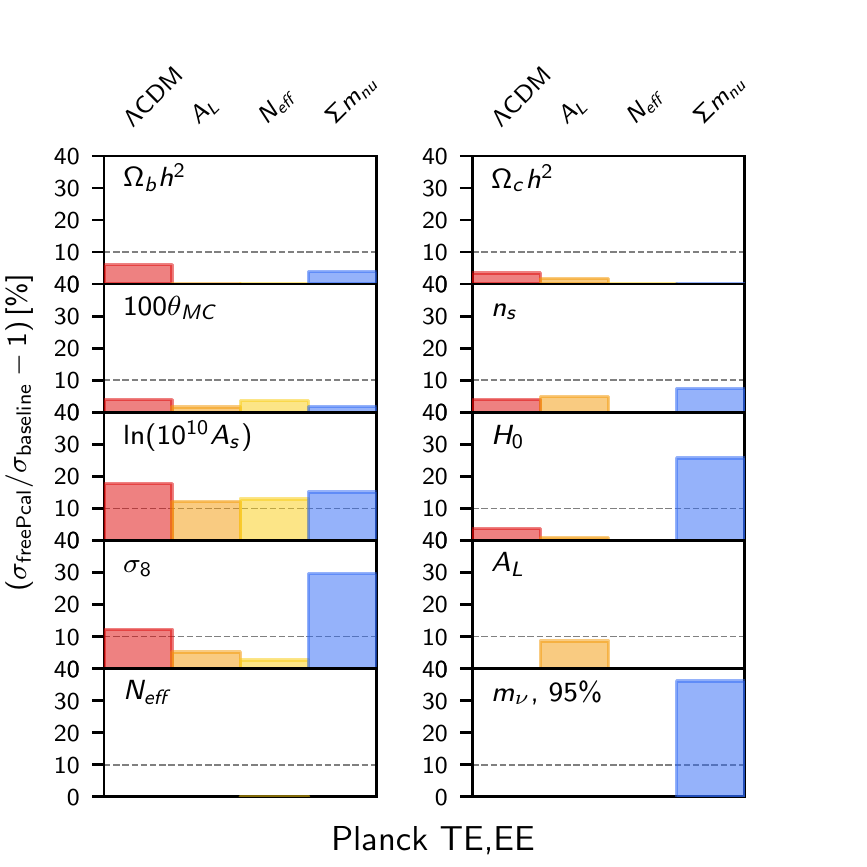}
\caption{Same as Figure~\ref{fig:sptpol_cosmo}, but for the \planck\ \TE,EE\ data.  Freeing the \planck\ \pcal\ parameters for this data combination has a large impact only in the $\Lambda$CDM+$\Sigma m_\nu$ case, where the 95\% confidence level upper limit on the sum of neutrino masses $\Sigma m_\nu$ and the error bars on derived parameters $H_0$ and $\sigma_8$ are increased by $30-40\%$. This is due to a shift in the best fit values of $\lnAs$, rather than an increase in degeneracies between parameters, see~\refsec{planckext}.}
\label{fig:planckcosmo}
\end{center}
\end{figure}

\begin{figure}
\begin{center}
\includegraphics[width=0.5\textwidth]{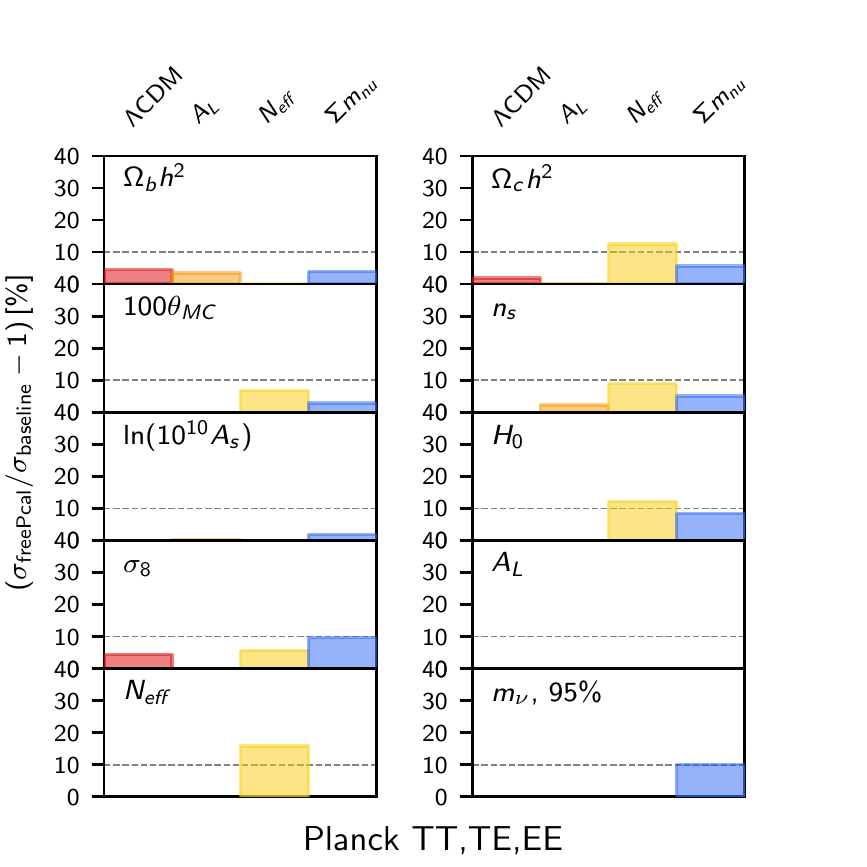}
\caption{Same as Figure~\ref{fig:sptpol_cosmo}, but for the \planck\ \TT,TE,EE\ data.  Freeing the three \planck\ \pcal\ frequency parameters has a very minor impact on the cosmological parameter error bars, smaller than $15\%$, in all the cosmological models considered here. }
\label{fig:planckcosmoTTTEEE}
\end{center}
\end{figure}

We now turn to extensions to the $\Lambda$CDM model.
Similar to~\refsec{sptpol}, we check the constraints on \pcal\ for three extensions, $\Alens$, $\mnu$, and $\nnu$.
The \pcal\ uncertainties are shown 
in~\reffig{planckpolTTTEEE} for TE,EE and TT,TE,EE.
We see that in all cases, the uncertainties of the \pcal\ parameters are similar to those in $\Lambda$CDM.
As for the cosmological parameter uncertainties, Figures~\ref{fig:planckcosmo} and~\ref{fig:planckcosmoTTTEEE} show the increase in their error bars when marginalizing over polarization calibration parameters for \planck\ TE,EE and TT,TE,EE respectively.
 
The parameter uncertainties in $\Lambda$CDM+$\nnu$ are little affected, with increases in the error bars by less than 15\%.
On the contrary, we find a somewhat larger effect on parameter uncertainties in the $\Lambda$CDM+$\mnu$ model for the \TE,\EE\ data. 
In this case, marginalizing over \pcal\ increases the upper limit on $\mnu$ by almost $40\%$, while degrading the uncertainties on $H_0$ and $\sigma_8$ by almost $30\%$. 
We note that the main source causing the degradation in $\mnu$ does not come from a drastic increase in posterior uncertainty given the degeneracy between $\mnu$ and \pcal.
The main effect rather comes from a shift in the best-fit values of correlated parameters $\mnu$, $\lnAs$, and \pcal. 
For this data set, \TE\ dominates the fit and causes $\mnu$ and $\lnAs$ to be anti-correlated. 
With \pcal\ free, the best fit for $\lnAs$ shifts to lower values by about $0.7~\sigma$. 
Thus, a lower value of $\lnAs$ induces a shift of the $\mnu$ posterior distribution to higher values. 
Since this distribution is single-tailed with $\mnu>0$, this shift is perceived as a change in the upper bounds.
These degradations disappear once the \TT\ data is included, because \TT\ strongly constrains $\lnAs$.
While this shift could be due to either a statistical fluctuation or a systematic error, it highlights the impact of \pcal\ on constraining $\mnu$.

For the $\Lambda$CDM+$\Alens$ model, it was noted in \planck\ that the $\Alens$ parameter is high compared to the $\Lambda$CDM expectation---at the 2.8\,$\sigma$ or 2.1\,$\sigma$ levels\footnote{These results refer to the baseline data combination \TT,\TE,\EE+\lowE+CMB lensing. Note that the $\Alens$ parameter only impacts the amplitude of lensing in the \TT,\TE,EE\ power spectra, while it leaves the \planck\ CMB lensing reconstruction power spectrum unaltered.} for polarization calibrations estimated using \plik's baseline or estimated using separate fits of \TE\ and \EE\ respectively, as already described above in \refsec{planckdata}. 
Here we show that leaving the polarization calibrations free to vary cannot alleviate the difference between these two results. This is due to the fact that the difference between the two \planck\ estimates of polarization efficiency from \TE\ alone or from \EE\ alone ($\Delta$\pcal$\sim0.017$ at 143 Ghz at map level) is larger than the \pcal\ posterior width when \pcal\ is free to vary when fitting the \TE,\EE\ or \TT,\TE,\EE\ data ($\sigma($\pcal$)\lesssim 0.01$). Furthermore, the  \pcal\ mean values measured from these fits are in good agreement with those of the baseline estimates.\footnote{This is not surprising since the \pcal\ fits obtained from the \TE,\EE\ or \TT,\TE,\EE\ data are dominated by \EE, which is also the data set used for the baseline estimates.} Therefore, leaving \pcal\ free to vary provides results which are similar to the baseline case.
Specifically, using the \TE,\EE+lowE data set, the $\Alens$ parameter best fit is $\Alens=1.09\pm 0.13$, 
which is within $0.8\,\sigma_{\rm exp}$ of the value obtained when fixing \pcal\ in the baseline case, $\Alens=1.13\pm 0.12$, with negligible impact on the uncertainties. 
Similarly when also including \TT, varying the polarization calibrations leads to $\Alens=1.19 \pm 0.069$, in agreement with the baseline result obtained with \pcal\ fixed $\Alens=1.18\pm 0.068$ (see also the discussion in section 3.7 of ~\cite{planck2018likelihood}).
Thus, leaving the polarization calibrations free to vary has a very small impact on the value and error bar of the $\Alens$ parameter, which remains higher than unity at the $2.8\sigma$ level, due to the tight constraint provided by the \TE,EE\ or \TT,TE,EE\ data combinations which agree with the baseline estimate. 
For the same reason, the other cosmological parameters are little affected as well.

\section{Forecasts}
\label{sec:forecasts}
In this section, we forecast how well \pcal\ could be measured with our method and the impact on cosmological parameter uncertainties when marginalizing over \pcal\ for ongoing and future experiments. 
We consider two experiment configurations:  SPT-3G, the third-generation camera currently installed on the South Pole Telescope~\cite{benson2014, bender2018}, and CMB-S4, a next-generation ground-based CMB experiment~\cite{cmbs4dsr}. 

\subsection{SPT-3G}
The SPT-3G receiver observes in three frequency bands 95, 150, and 220\,GHz in both intensity and polarization with $\sim16000$ detectors
over $\sim$1500 deg$^2$ of the sky in its main survey field.
The full-width half-maximum of the beams are approximately 1.7, 1.2, and 1.1\,arcminutes at 95, 150, and 220\,GHz respectively.
The first science results from SPT-3G using \TE\ and \EE\ spectra measured using
data collected in 2018 have recently been released~\cite{eete2018}.
However, the data were only collected for half of the observing season with part of the focal plane operable.
Therefore, for this forecast, we use noise level projections starting from 2019 when the active detector count nearly doubled.
With five seasons of observations on the main survey field (2019--2023 inclusive), the noise levels in the final coadded temperature maps are projected to be 3.0, 2.2, and 8.8\,$\mu$K\,arcmin in the three frequency bands, and those in the polarization maps are a factor of $\sqrt{2}$ higher~\cite{benson2014, bender2018}.
 
We forecast the \pcal\ constraints along with constraints on $\Lambda$CDM and extension parameters for SPT-3G for two scenarios. 
First, we use data from only one of the three frequency bands, $150$GHz, for more direct comparison with \sptpol, described in~\refsec{sptpol}, and to verify the impact of using only one frequency channel.
Second, we report the constraints when combining maps from all three bands. 

We use the Fisher Matrix formalism and code described in~\cite{galli2014} for extracting the 1-$\sigma$ parameter uncertainties. 
As inputs, we use lensed power spectra of \TT, \TE, and \EE; we do not include the lensing reconstruction spectrum $C_L^{\phi\phi}$. 
We present constraints from the combination of \TE\ and \EE\ as a baseline and also those including all three spectra to study the effect of including \TT. 
We restrict the power spectrum angular multipole range to $\ell=100-3500$, and we adopt a Gaussian prior on the optical depth to reionization of $\sigma(\tau)=0.007$, based on the Planck constraint~\cite{planck18cosmoparam}. 
We check that including $1/f$ noise or marginalizing over foregrounds do not change these results substantially. 

Table~\ref{tab:forecast} shows results for the $\Lambda$CDM case. 
The SPT-3G \TE\ and \EE\ combination is projected to constrain \pcal\ at the level of $\sim 0.8\%$, either using only one frequency or combining the information from all three frequencies. 
When freeing \pcal, the constraint on $\lnAs$ is degraded by about $50\%$ while the rest of the $\Lambda$CDM parameters are mildly affected (below the $15\%$ level). 
Similar to what is seen in the \planck\ case, the degraded constraints can be recovered by adding the \TT\ data. 
In this case, marginalizing over \pcal\ has negligible impact on cosmological parameters and the constraint on \pcal\ tightens to $0.2\%$. 

We verify that similar constraints on \pcal\ are obtained in extensions of the $\Lambda$CDM model, such as $\Lambda$CDM+$\nnu$, $\Lambda$CDM+$\Alens$ or $\Lambda$CDM+$\mnu$, for both the \TE+\EE\ and the \TT+TE+EE\ data combination. 
As for the cosmological parameters, we highlight here the ones with constraints degraded when marginalizing over \pcal. 
In $\Lambda$CDM+$\mnu$, the $\lnAs$ uncertainty increases by 40\% for the \TE+EE\ data combination. 
In $\Lambda$CDM+$\nnu$, the $\lnAs$ uncertainty increases by 70\% and the uncertainties on $\Omega_b h^2$ and $H_0$ increase by $\sim 30\%$. 
However, similar to the $\Lambda$CDM case, when including the \TT\ data, the marginalization over \pcal\ has minimal impact on the constraints on cosmological parameters.

\begin{table*}[ht!]
\centering
\caption{Fisher matrix forecast on cosmological parameters and \pcal\ for SPT-3G, using the 150\,GHz channel alone or all of the three channels. As a comparison, we also show constraints when fixing \pcal.}
\label{tab:forecast}
\begin{tabular}{l|c|c|c|c|c|c|c}
\hline
\hline
&$\Omega_b h^2$ & $\Omega_c h^2$ & $H_0$ & $\tau$ & $ns$ & $ln[10^{10} As]$ & $P_\mathrm{cal}$ \\
&$[\times 10^{-4}]$ &$[\times 10^{-3}]$&$[\times 10^{-1}]$&$[\times 10^{-3}]$&$[\times 10^{-3}]$&$[\times 10^{-2}]$&$[\times 10^{-3}]$\\
\hline
$\Lambda$CDM\\\hline
SPT-3G \TE+\EE\, 150GHz &1.4 &    2.0  &   7.5 &    6.6 &    8.0  &   1.3\\

SPT-3G \TE+\EE\ & 1.3  &   1.9  &   7.1   &  6.6  &   7.7&     1.3\\
SPT-3G \TT+\TE+\EE\ & 1.4&1.7&     6.5&     6.4&     7.4&     1.2 \\
\hline
$\Lambda$CDM+$P_{cal}$\\\hline
SPT-3G \TE+\EE\ 150GHz & 1.6    & 2.1 &    8.0   &  6.6&     8.2  &   2.0 &    7.6 \\

SPT-3G \TE+\EE\ &1.5   &  2.0  &   7.7 &    6.6     & 7.9    & 1.9   &  7.4\\
SPT-3G \TT+\TE+\EE\ &1.4&     1.8&     6.8&    6.4&     7.4&     1.2&     2.1\\
\hline
\hline
\end{tabular}
\end{table*}

\subsection{CMB-S4}

CMB-S4 is a next-generation ground-based CMB experiment aiming to observe $\sim$70\% of the sky.
It is planned to have a frequency coverage from 20 to 270\,GHz and the full-width half-maximum of its beam at 150\,GHz is $\lesssim$1.5 arcminutes~\cite{cmbs4dsr}.
There will be telescopes observing from both the South Pole and from the Atacama desert in Chile, for a deep and a wide area survey respectively. 

In this work, we forecast the constraints on \pcal\ given the wide survey from Chile. 
We use noise curves from \cite{hill}, which combine information from all frequencies using an internal linear combination method. 
The per-frequency noise input includes atmospheric noise; the output noise curves include residuals from component separation. 
We assume $f_{sky}=0.42$, which excludes the area covering the galaxy in the wide survey. 
As in the forecast for SPT-3G, we use lensed power spectra in the multipole range of $\ell=100-3500$ and we do not include information from lensing reconstruction $C_L^{\phi\phi}$. 

Table~\ref{tab:forecast2} shows results for the $\Lambda$CDM case. 
We find that with just \TE\ and \EE, CMB-S4 data could constrain \pcal\ at the level of $\sim 0.2\%$, which further tightens to $0.056\%$ when we add \TT. 
When freeing \pcal, constraints on cosmological parameters are mildly degraded without \TT, and negligibly degraded with \TT. 
As in the previous sections, we verify that extending the $\Lambda$CDM model with $\mnu$, $\nnu$, and $\Alens$ 
does not significantly change the constraints on \pcal.
Conversely, leaving the \pcal\ parameter free has the largest impact on the constraints on $\Omega_b h^2$, $H_0$ and $\lnAs$ in the $\Lambda$CDM+$\nnu$ model for \TE+EE, at the level of 30\%. 
Similarly to previous cases, including the \TT\ data allows us to marginalize over \pcal\ with no loss of precision on cosmological parameters.

\begin{table*}[ht!]
\centering
\caption{Fisher matrix forecast on cosmological parameters and \pcal\ for CMB-S4. As a comparison, we also show constraints when not varying the \pcal.}
\label{tab:forecast2}

\label{tab:forecast2}
\begin{tabular}{l|c|c|c|c|c|c|c}
\hline
\hline
&$\Omega_b h^2$ & $\Omega_c h^2$ & $H_0$ & $\tau$ & $ns$ & $ln[10^{10} As]$ & $P_\mathrm{cal}$ \\
&$[\times 10^{-4}]$ &$[\times 10^{-3}]$&$[\times 10^{-1}]$&$[\times 10^{-3}]$&$[\times 10^{-3}]$&$[\times 10^{-2}]$&$[\times 10^{-3}]$\\
\hline
$\Lambda$CDM\\\hline
CMB-S4 \TE+\EE\ & 0.36  &   0.71  &   2.7  &   5.1&     2.5  &   0.88\\
CMB-S4 \TT+\TE+\EE\ &0.36   &  0.67  &   2.5  &   4.9 &    2.3 &    0.85\\
\hline
$\Lambda$CDM+$P_{cal}$\\\hline
CMB-S4 \TE+\EE\ & 0.42 & 0.75  &   2.9 &    5.1 & 2.5 & 1.0 &     2.0 \\
CMB-S4 \TT+\TE+\EE\ & 0.37 &     0.70 &  2.6 &     4.9 &     2.3 &     0.86 &     0.56\\
\hline
\hline
\end{tabular}
\end{table*}

\section{Conclusions}
\label{sec:conclusions}
In this paper, we demonstrate that effective polarization calibrations \pcal\ could be precisely determined by 
fitting CMB \TE\ and \EE\ spectra to the $\Lambda$CDM model and its common extensions with \pcal\ as a free parameter. 
This is possible thanks to the different dependence of the \TE\ and \EE\ spectra on \pcal. 
While allowing \pcal\ to float does increase the posterior volume and therefore degrades some constraints on
cosmological parameters, we show that the degradation becomes negligible once \TT\ is included. 

We apply the method to \sptpol\ and \planck. 
For the \sptpol\ 150\,GHz \TE\ and \EE\ data set presented in H18, 
we extract \pcal\ with an uncertainty of $\sim 2\%$ at the map level, independent of the considered models. 
For the data set from the \planck\ 2018 data release, combining \TE\ and \EE\ allows us to measure \pcal\ at 100, 143, and 217\,GHz with uncertainties of $0.7\%$, $0.6\%$ and $0.8\%$ at the map level.
We highlight how this method can be useful for detecting inconsistencies in the data. 
In particular, \pcal\ determined using \TE\ and \EE\  should agree with the ones determined with \TT\ included or the ones measured from external data sets. 
If not, this could suggest the existence of unaccounted for systematics which project into these multiplicative factors.

Finally, we forecast the capabilities of current and future experiments to constrain \pcal.  
We find that using its 3 frequency channels, SPT-3G will be able to measure \pcal\ with an uncertainty of 0.7\% from \TE\ and \EE, 
and the uncertainty can be improved to 0.2\% when including \TT. 
We find that leaving \pcal\ free to vary will degrade the constraints on $\As$ from \TE\ and \EE\ by 30\%, while constraints from \TT,\TE,\EE\ are not affected.
Furthermore, we find that CMB-S4 could further tighten the uncertainty of \pcal\ to 0.2\% with its \TE\ and \EE\ measurements and to 0.06\% with \TT,\TE,\EE. 
Similarly to SPT-3G, while constraints on $\As$ are affected by the variation of \pcal\ by about 20\% when using \TE,\EE, the constraints from \TT,\TE,\EE\ are unaffected.

We highlight that these uncertainties on \pcal\ are comparable to or tighter than those derived for the  \planck\ baseline ($\sim0.5\%$). 
As a consequence, relying on \planck\ to calibrate polarization maps will ultimately limit the accuracy of these experiments, provided that the \planck\ uncertainty is folded in the power spectrum covariance matrix. 
Furthermore, if the external \pcal\ determination is biased and the systematic uncertainties are not properly included, cosmological parameters constraints would be biased. 
We observe a possible hint of this in the \planck\ \TE,\EE\ $\mnu$ upper limits between the (baseline) fixed \pcal\ case and the free \pcal\ case. For \planck\ however, the difference of $\mnu$ upper limits due to a shift in the \pcal\ values is still compatible with a statistical fluctuation. For future experiments, we emphasize that stringent control on \pcal\ is important
for accurate and precise inference on cosmological parameters, such as the sum of neutrino masses. 

Beyond using the primary CMB spectra \TT, \TE, and \EE, we acknowledge the possibility of adding lensing potential power spectrum measurements to further tighten constraints on \pcal\ and reduce degradations in cosmological parameters. We leave this for future work.
In conclusion, this paper demonstrates that a significant source of systematic error for future CMB polarization experiments can be self-calibrated without major consequences on the constraints on cosmological parameters.

\begin{acknowledgements}
We thank the participants of the CMB systematics and calibration focus workshop hosted virtually by Kavli IPMU for helpful comments and feedback.
This work was completed in part with resources provided by the University of Chicago's Research Computing Center.
This work has received funding from the French Centre National d'Etudes Spatiales (CNES). This research used resources of the IN2P3 Computer Center (\url{http://cc.in2p3.fr}). 
WLKW is supported in part by the Kavli Institute for Cosmological Physics at the University of Chicago through grant NSF PHY-1125897, an endowment from the Kavli Foundation and its founder Fred Kavli, and by the Department of Energy, Laboratory Directed Research and Development program and as part of the Panofsky Fellowship program at SLAC National Accelerator Laboratory, under contract DE-AC02-76SF00515. 
TC is supported by the the National Science Foundation through South Pole Telescope grant OPP-1852617.
\end{acknowledgements}  

\bibliography{pcal}
\end{document}